\documentclass[a4paper,fleqn,usenatbib]{mnras}

\usepackage[T1]{fontenc}
\usepackage{ae,aecompl}

\usepackage{graphicx}	% Including figure files
\usepackage{amsmath}	% Advanced maths commands
\usepackage{amssymb}	% Extra maths symbols
\usepackage[none]{hyphenat}
\usepackage[T1]{fontenc} 
\usepackage{aecompl}

\title[]{Kinematics of the Galactic SNR G109.1-1.0 (CTB\,109)}

\author[M. S\'anchez-Cruces et al.]{
M. S\'anchez-Cruces,$^{1}$\thanks{E-mail: msanchez@esfm.ipn.mx}
M. Rosado$^{2}$,
I. Fuentes-Carrera$^{1}$
and P. Ambrocio-Cruz$^{3}$
\\
$^{1}$Escuela Superior de F\'isica y Matem\'aticas, Instituto Polit\'ecnico Nacional, U.P. Adolfo L\'opez Mateos, C.P. 07738,\\
Ciudad de M\'exico,\\
$^{2}$Instituto de Astronom\'ia, Universidad Nacional Aut\'onoma de M\'exico, Circuito Exterior, C.U., A. Postal 70-264, 04510 \\
Ciudad de M\'exico, M\'exico\\
$^{3}$Escuela Superior de Tlahuelilpan, Universidad Aut\'onoma del Estado de Hidalgo, Ex-Hacienda de San Servando s/n, Col. Centro,\\
 42780 Tlahuelilpan Hgo., M\'exico
}

\date{Accepted XXX. Received YYY; in original form ZZZ}

\pubyear{2017}

\begin{document}
\label{firstpage}
\pagerange{\pageref{firstpage}--\pageref{lastpage}}
\maketitle

\begin{abstract}
We present direct images in the H$\alpha$ and [SII]$\lambda \lambda$6717,6731 $\text{\AA}$ lines of the Galactic Supernova Remnant G109.1-1.0 (CTB 109). We confirm that the filaments detected are the optical counterpart of the X-ray and radio supernova remnant due to their high [SII]/H$\alpha$ line-ratios. We study for the first time the kinematics of the optical counterpart of SNR CTB\,109 using the UNAM scanning Fabry-Perot interferometer PUMA. We estimate a systemic velocity of V$_{LSR}$=-50$\pm$6~km~s$^{-1}$ for this remnant and an expansion velocity of V$_{exp}$=230$\pm$5~km~s$^{-1}$. From this velocity value and taking into account previous studies about the kinematics of objects at that Galactic longitude we derive a distance to the SNR CTB\,109 of 3.1$\pm$0.2 kpc, locating it in the Perseus arm. Using the [SII]$\lambda$6717/[SII]$\lambda$6731 line-ratio we find an electronic density value around n$_e$= 580~cm$^{-3}$.
Considering that this remnant is evolving in a low density medium with higher density cloudlets responsible of the optical emission, we determine the age and  energy deposited in the ISM by the supernova explosion (E$_0$) in both the Sedov-Taylor phase and the radiative phase. For both cases the age is of thousands of years and the E$_0$ is rather typical of SNRs containing simple pulsars so that, the energy released to the ISM cannot be used to distinguish between supernova remnants hosting typical pulsars from those hosting powerful magnetars as in the case of CTB\,109.

\end{abstract}


%%%%%%%%%%%%%%%%%%%%%%%%%%%%%%%%%%%%%%%%%%%%%%%%%%
%%%%%%%%%%%%%%%%% BODY OF PAPER %%%%%%%%%%%%%%%%%%
%\hspace{3cm}

\section{Introduction}
A Supernova Remnant (SNR) is the region confined by the shock wave generated by a supernova explosion. It contains both shocked interstellar medium and the material ejected   by the explosion. Depending on the mass of the progenitor star and the circumstellar environment at the moment of the explosion (core-collapse of massive stars or thermonuclear explosion in a binary system) a compact object can remain. The energy deposited in the interstellar medium (ISM) is of the order of 10$^{50}$ to 10$^{51}$~erg.

The supernova remnant G109.1-1.0 (CTB 109) is located at $\alpha_{(J2000)}$=23$^h$01$^m$35$^s$, $\delta_{(J2000)}$=+58$^{\circ}$53'
according to its source centroid  \citep[Green Catalogue,][]{Green2014}. Its morphology shows a semicircular shell both in radio emission \citep{Hughes1981} and X-ray \citep{Gregory-Fahlman1980} where it has been discovered. X-ray emission from ROSAT HRI \citep{Hurford-Fesen1995}, XMM-Newton \citep{Sasaki2004} and Chandra \citep{Sasaki2013} reveal the shell-type and semicircular morphology of the SNR CTB\,109. 

\begin{figure*}
\includegraphics[width=2\columnwidth]{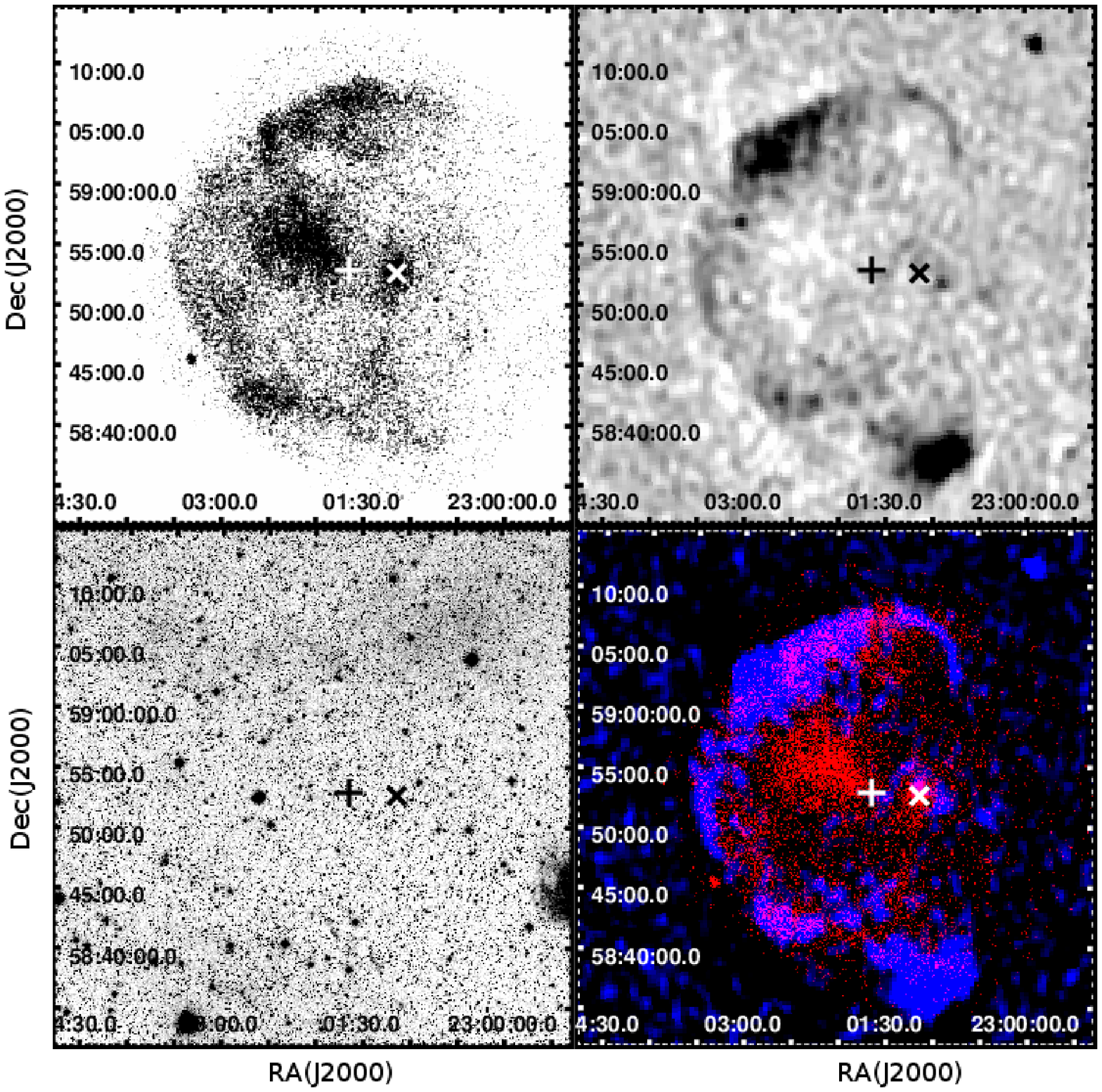}
  \caption{\textit{Top-left}:X-ray image of the SNR CTB\,109 taken from ROSAT HRI \citep{Hurford-Fesen1995}. \textit{Top-right}: Radio continuum image (at 325 MHz (92 cm)) taken from The Westerbork Northern Sky Survey \citep{Rengelink1997}. \textit{Bottom-left}: Optical image taken from DSS data base. \textit{Bottom-right}: Composed image of X-ray (red) and radio continuum (blue). ``$\times$'' indicates the location of the X-ray magnetar AXP~1E~2259+586 and ``+'' is the geometrical centre of the almost hemispherical shape emitting gas \citep{Kothes2006}.}
  \label{F1}
\end{figure*}

In the Sloan Surveys there is no appreciable optical emission in the location of the X-ray emitting hemispherical shell (see Figure \ref{F1}). Optical emission of this remnant was detected long time after the X-ray and radio emissions by \cite{Fesen-Hurford1995} who found H$\alpha$ filaments whose spectra showed [SII]/H$\alpha$ ratios consistent with shocks. 
Regarding CO and HI studies, none of them was successful in detecting shocked molecular or neutral material directly associated with the SNR. Nevertheless, the molecular gas observations show an anticorrelation between the X-ray hemispherical shell and the CO clouds suggesting a possible interaction \citep[see][]{Tatematsu1987, Kothes2002, Sasaki2006, Tian2010, Kothes-Foster2012}.

In the ROSAT and XMM-Newton images it is possible to see the  anomalous X-ray pulsar (AXP) 1E~2259+586. This pulsar is also known as MG J2301+5852 in the McGill magnetar catalogue \citep{Olausen-Kaspi2014} and as AXP~J2301+5852.  This central source was discovered by \cite{Gregory-Fahlman1980} as an X-ray point source. One year later the central source was reported as an X-ray pulsar \citep{Fahlman-Gregory1981}. Now this source is considered a magnetar \citep{Kaspi2002, Gavriil2004, Woods2004} with a  magnetic field of 5.9$\times$10$^{13}$~G \citep{Tendulkar2013}. The measured period is  6.97~s \citep{Morini1988}. No radio, nor optical emission has been detected for the magnetar.

The difference between a magnetar and a pulsar lies on the values of the period and the surface magnetic field. Typically the magnetic field of a magnetar is about 10$^{13}$-10$^{15}$~G \citep{Woods-Thompson2006, Mereghetti2008, Harding2013} while the magnetic field of a pulsar is about 10$^{12}$~G \citep{Popov2010}. The period of a magnetar is >1 s while that of a pulsar is less or about 1~s. 

Figure~\ref{F1} shows that the SNR CTB\,109 presents strong correlation between X-ray emission and radio emission. No optical emission is detected in the DSS\footnote[1] {SkyView Survey metadata. Data taken by ROE and AAO, CalTech, Compression and distribution by Space Telescope Science Institute.} image at that level associated with the SNR. The X-ray magnetar AXP~1E~2259+586 (marked with a ``$\times$'') and the geometric centre of the remnant given by \cite{Kothes2006} (marked with ``+'') do not match. If we assumed that the SNR  expanded symmetrically, this mismatch could suggest a transverse motion of the magnetar.

\subsection{Previous distance estimates}

In this section we present a brief summary of the previous distance determinations to the SNR CTB\,109 and to the magnetar AXP 1E 2259+586.

Given that the SNR CTB\,109 is in the direction of the Perseus arm, distance estimates based on the kinematics should be checked  with other distance indicators. Large differences between the radial velocities and the distance estimated by others means than the kinematical one have been detected in the direction of the Perseus arm \citep{Georgelin-Georgelin1976, Fich1989, Brand-Blitz1993}. These so called ``velocity anomalies'' make the systemic velocities more negative and thus, they tend to overestimate the kinematic distance.

The ``velocity anomaly'' is thought to be due to a slowing down of the interstellar matter of the Perseus arm caused by the spiral shock developed in the arm. This results in  a more negative radial velocity for HII regions, supernova remnants and planetary nebulae in the arm.

Based on the density wave and Galactic-shock wave concepts proposed for the Perseus arm, \cite{Roberts1972} used the two-armed spiral shock model (TASS) to describe the large scale motion of the interstellar gas in the Galactic disk and to explain  this ``velocity anomaly''. In that model the Perseus arm is consistent with a Galactic shock wave located at about 2 to 3 kpc (see his Figure 3). In that sense, the ``velocity anomaly'' is a large scale motion due to the shocks and spiral arms. \cite{Roberts1972} identifies different zones in a velocity-distance diagram that describes the kinematics in the direction of the Perseus arm: a shock jump, a region of dense gas behind the shock, a velocity ``hill'', and a velocity trough on the location of a local gas density peak.

\cite{Foster-MacWilliams2006} developed another method to deal with distances in the direction of the Perseus arm. The method uses an empirical fitting of HI fluxes in order to find distances along the Perseus arm direction.  This method is not based in the hydrodynamics equations used by \cite{Roberts1972} to describe the shock. Nevertheless, their results are similar to those obtained by \cite{Roberts1972}. 

Regarding the  distance determinations to the SNR CTB\,109, \cite{Kothes2002} used distances derived from stellar spectra of 16 exciting stars of  HII regions near SNR CTB\,109  and compared them with the kinematic distances of the same HII regions derived from HI and CO observations from the Canadian Galactic Plane Survey \citep[CGPS,][]{Taylor2003}. Their Figure 6 indicates that at these Galactic longitudes, the distances to the HII regions are between 3 to 4 kpc using the \cite{Taylor-Cordes1993} determination. For the SNR CTB\,109 there is no conspicuous HI or CO emission which could be used to show that it is the counterpart of that SNR. As mentioned before, \cite{Tatematsu1987} have found an anticorrelation between the hemispherical shape revealed by the SNR emission both in radio continuum and in X-rays and the CO. This anticorrelation is taken as proof that SNR CTB\,109 is interacting in its western side with the molecular cloud, however no shocked CO has been detected.  \cite{Kothes2002} measured the velocities of CO and HI in the direction of the SNR. These authors found that the radial velocity in the direction of the SNR CTB\,109 is between -50 and -52 km s$^{-1}$ for the CO and between -47 and -53~km~s$^{-1}$ for the HI. These velocities agree with those of the HII regions derived previously. According to the two-armed spiral shock model of \cite{Roberts1972} they lie near the position of the spiral shock in the Perseus arm. 

However, there is still a distance ambiguity in order to locate the SNR CTB\,109 using the \cite{Roberts1972} model: at the shock jump and at the velocity hill. \cite{Kothes2002}  assumed that a SN explosion generating a pulsar (or magnetar) should come from a massive star that does not have enough time to go far away from its birth place. Thus they place the SNR at the shock jump at 3$\pm$0.5~kpc. They support this argument by comparing emission and absorption HI velocity profiles in the direction of the SNR and also the HI profiles of two nearby HII regions whose stellar distances place them in the Perseus arm and an extragalactic source, respectively. 

\cite{Durant-vanKerkwijk2006} estimated the distance to the magnetar AXP~1E~2259+586 using the ``red clump method'' to provide an interstellar distance-extinction relationship for several anomalous X-rays pulsars or magnetars including  AXP~1E~2259+586.
The ``red clump method'' relies on the identification in an IR color-magnitude diagram of a conspicuous strip to the right of the main sequence presumably formed by core He-burning giants which act as standard candles. The differences (and absences) along that strip are interpreted as differences in reddening N$_H$ and A$_v$.  They made an extrapolation on the visual extinction A$_v$ with the X-ray extinction estimates for the AXPs calculated with the absorbing total hydrogen column density N$_H$. For AXP~1E~2259+586 they estimated an A$_v$ of about 6.4 mag (a very high value) and thus, they derived a distance of 7.5$\pm$1.0~kpc. However, as pointed out \cite{Kothes-Foster2012} and in a previous work by \cite{Vink2008}, this method should be considered with care. These authors list some problems of the \cite{Durant-vanKerkwijk2006} method: 1) The large intrinsic spread in the N$_{H}$-A$_v$ relation. 2) The method does not consider the discrepancy between the large gradient of the absorbing HI column density in the direction of CTB\,109 and the N$_{H}$ of the magnetar AXP~1E~2259+586. 3) The comparison of the foreground extinction A$_v$ determined over
a very large angular extent with that of a pulsar, which is a punctual source, without taking into account the large spread of visual extinction values in that region going from practically zero to up to 6.5 mag, and thus giving large uncertainties in the distance.

\cite{Tian2010} used the HI line and CO-line observations from the CGPS already used by \cite{Kothes2002}. They compared the absorption line profiles in the direction of the SNR CTB\,109, with the profiles of the HII region Sh\,2-152 and the adjacent molecular cloud complex. They adopt the velocity of the molecular complex of -55~km~s$^{-1}$ as the velocity of the SNR CTB\,109.
They  use the \cite{Foster-MacWilliams2006} method to get the distance to the SNR. They chose the farthest distance to the SNR of 4 kpc using as additional information the distance value given by  the method of \cite{Roman-Duval2009} to solve the kinematics ambiguity. This method is based on testing whether certain molecular clouds are related to observed HI self-absorption (HISA) features or not. However, \cite{Kothes-Foster2012} attributed the non detection of HISA in \cite{Tian2010}  due to the fact that the absorption amplitude detected in the direction of CTB\,109 is below  the cutoff used for the automatic detection program in \cite{Gibson2005}. 

In their 2012 work, Kothes \& Foster analysed the three distance determinations presented above \citep{Kothes-Foster2012}. They also estimated a kinematic distance to SNR CTB\,109 based on the possible interaction of the SNR with the molecular cloud studied by \cite{Tatematsu1987}. \cite{Kothes-Foster2012} used the previous velocity range of the molecular and neutral material in the direction of SNR CTB\,109  (from -48 to -56~km~s$^{-1}$). They also used the HI velocity profiles obtained with the CGPS to fit them by means of the \cite{Foster-MacWilliams2006} method. They obtained three possible distance values. The farther distance lies between 3.9 and 4.3 kpc placing the SNR in the interarm region. The intermediate distance places the SNR slightly downstream of the shock, at 3.2$\pm$0.2 kpc. Their closest value places the SNR at the Perseus shock jump (3.0$\pm$0.2 kpc).

\cite{Kothes-Foster2012} eliminate the far distance value by analysing HI absorption profiles in the direction of the SNR and those corresponding to two nearby HII regions and another profile of an extragalactic source. These authors also reasoned that the progenitor of the SNR CTB\,109 should be a massive star formed in the shock which before exploding as a supernova it migrated to a position beyond the shock but still in the Perseus arm because no HI shell of a very massive star is detected.  By comparing the progenitor longevity with those stars earlier than $\sim$B2, those authors find that the progenitor could have migrated from the jump shock to the velocity hill (200 pc) in order to reach a distance of 3.2$\pm$0.2 kpc. Therefore they conclude that the SNR CTB\,109 is located at that distance.

\begin{table}\centering
	\caption{Characteristics of CTB\,109}
    \label{T1}
    \scalebox{0.9}{
    \begin{tabular}{lcc} 
		\hline
 Parameter & 
 Value&
 Reference\\
 \hline
Galactic 	&  \textit{l}=109.1$^{\circ}$, \textit{b}=-1.0$^{\circ}$ 	& \cite{Green2014} \\
coordinates\\
Equatorial  	& $\alpha_{J2000}$=23:01:35 			& \cite{Green2014} \\
coordinates		& $\delta_{J2000}$=+58:53              &  \\
Distance 				& 3.2 kpc		    & \cite{Kothes-Foster2012}\\
     					& 4.0 kpc 			& \cite{Tian2010}\\
     					& 3.0 kpc   	    & \cite{Kothes2002} \\
     					& 7.5 kpc$^{a}$  	   	 & \cite{Durant-vanKerkwijk2006}\\
     					& 2.8 kpc $^{b}$           & \cite{Cordes-Lazio2002}\\
Age						& 1.9$\times$10$^{4}$~yr 	&\cite{Rho-Petre1997}\\
						& 1.6$\times$10$^{4}$~yr$^{c}$ 	&\\
						& 8.8 $\times$10$^{3}$~yr	& \cite{Sasaki2004}	\\
						& 1.4$\times$10$^{4}$~yr	& \cite{Sasaki2013}	\\
						& 1.3$\times$10$^{4}$~yr$^{c}$ 	&\\
						& 1.4$\times$10$^{4}$~yr	& \cite{Nakano2015}	\\
						& 1.3$\times$10$^{4}$~yr$^{c}$ 	&\\
						& 2.4$\times$10$^{4}$~yr	& \cite{Tendulkar2013}	\\
\hline	
\end{tabular}} \\ 
\textit{a}~Distance to the magnetar AXP~1E~2259+586.\\
\textit{b}~By using the DM=79 pc cm$^{-3}$ given by \cite{Malofeev2006}.\\
\textit{c}~Considering the distance of 3.1 kpc.\\
\end{table}

\subsection{Age estimates}

In this section we present previous age estimates of the SNR CTB\,109 and the magnetar AXP~1E~2259+586 reported by different authors. 

Most of the studies used X-ray observations and considered the Sedov-Taylor similarity solution \citep{Sedov1959, Taylor1950} to obtain the age of the SNR CTB\,109. For example, \cite{Rho-Petre1997}  used ROSAT Position Sensitive Proportional counter (PSPC) and Broad Band X-Ray Telescope (BBXRT) data and fitted the spectrum of the northern shell with a two-temperature Raymond-Smith model. They considered a distance to the remnant of 4 kpc and used the Sedov-Taylor similarity solution to derive a shock velocity of v$_x$=380~km~s$^{-1}$, an electron density of n$_x$=1.2~cm$^{-3}$ and a remnant age of 1.9$\times$10$^{4}$ yr (1.6$\times$10$^{4}$ yr, considering the distance of 3.1 kpc). They inferred an initial energy explosion of E$_0$ of 1.5$\times$10$^{51}$ erg.

\cite{Sasaki2004} used the X-ray Multimirror Mission (XMM-Newton) European Photon Imaging Camera (EPIC) data of SNR CTB\,109. These authors assumed a distance to the SNR of 3.0 kpc estimated by \cite{Kothes2002} and  fitted the spectra of the eastern shell of the SNR. Using the Sedov-Taylor similarity solution, they obtained a shock velocity of v$_x$=720$\pm$60~km~s$^{-1}$, an age of 8.8$\pm$0.9$\times$10$^{3}$ yr, a preshock electronic density of n$_x$=0.16~cm$^{-3}$ and an initial energy E$_0$=7.4$\pm$2.9 $\times$10$^{50}$ erg. 

\cite{Sasaki2013} used the X-ray data of the SNR CTB\,109 from ACIS-I/Chandra.
They extracted X-ray spectra of 58 regions covering the northeastern  part of the SNR CTB\,109. They found that in general  all regions are fitted  with two thermal emission components, one to model the emission of the shocked ISM and the second to verify the existence of emission from shocked ejecta. Taking the distance of 3.2$\pm$0.2 kpc  estimated by \cite{Kothes-Foster2012} and using the Sedov-Taylor similarity solution, these authors found a blast wave velocity of 460$\pm$30 km s$^{-1}$ and an age of 1.4$\times$10$^{4}$~yr (1.3$\times$10$^{4}$ yr, considering the distance of 3.1 kpc).

\cite{Nakano2015} used X-ray Imaging Spectrometer (XIS) Suzaku observations to determine the age of the SNR CTB\,109 and its magnetar AXP~1E~2259+586. They fitted the eastern parts of the SNR assuming non-thermal equilibrium ionization. Taking the distance of 3.2$\pm$0.2~kpc  estimated by \cite{Kothes-Foster2012} and from the angular size of the SNR ($\sim$16') they estimated a radius R=16$\pm$1~pc.  Applying the Sedov-Taylor similarity solution, they calculated  a shock velocity of 460~km~s$^{-1}$, a preshock electronic density  n$_x$=0.1-0.3~cm$^{-3}$  and an age of 1.4$\times$10$^{4}$~yr (1.3$\times$10$^{4}$ yr, considering the distance of 3.1 kpc).

On the other hand, \cite{Tendulkar2013} used high resolution Near Infrared Camera (NIR) images from the 10m Keck 2 telescope to find the proper motion of AXP~1E~2259+586. They found values of $\mu_{\alpha}$=-6.4$\pm$0.6 mas yr$^{-1}$ and $\mu_{\delta}$=-2.3$\pm$0.6 mas yr$^{-1}$.
Then they considered the centre of the SNR CTB\,109 given by \cite{Kothes2006} $\alpha_{J2000}$=23h01m39s, $\delta_{J2000}$=+58$^{\circ}$53'00'' in order to compare the separation between the centre of the SNR CTB\,109 and the position of the magnetar  AXP~1E~2259+586 finding a kinematic age of 2.4$\times$10$^{4}$~yr. It is interesting to note that this age estimate does not depend on the distance assumed to the SNR. This age is larger than other estimated ages of the SNR CTB\,109 and those authors conclude that this could be due to a motion to the east of the centre of CTB\,109 due to the interaction to the west with a molecular cloud, whereas the magnetar moves to the west. We think that this larger value could also be due to the fact that the authors removed the ``Galactic rotation'' proper motion assuming that the progenitor of the magnetar  AXP~1E~2259+586 has been moving with the Galactic rotation curve without much dispersion. However, as discussed before, the objects in the Perseus arm are subject to a large scale motion (also called ``velocity anomalies'') produced by a shock in the spiral arm that slows down the motion of the objects in the arm relative to typical circular velocities. Thus, it is possible that this correction introduces some errors in evaluating the magnetar's age. 

All these age estimates are very consistent with each other except the  \cite{Tendulkar2013} value. But, even those authors conclude that the SNR and the magnetar are younger than their estimates and that it is more likely that the geometric centre is being displaced to the east. 

Table~\ref{T1} shows some properties of this remnant derived from the different previous studies mentioned here.

The layout of the paper is as follows: In Section 2, we present the observations and the data reduction, in Section 3, we present the derived kinematic parameters obtained for the SNR. Conclusions are presented in Section~4.

\begin{table}
	\centering
	\caption{Observational and Instrumental Parameters} 
\label{T2}
\begin{tabular}{lcc}
\hline
Parameter &
Value\\
\hline  
    Telescope							& 2.1~m (OAN, SPM) \\
    Instrument							& PUMA \\
    Detector							&  Marconi CCD\\
    Scale plate 						& 0.33''/pix \\
    Binning 							& 4 \\
    Detector size (pixels)				& 2048 $\times$ 2048 \\
    Cube dimensions						& 512 $\times$ 512 $\times$ 48\\
    Central Lambda (\AA) 			&  6720 \\
    Bandwidth (\AA) 				&  20\\
    Interference Order  		&  322 at 6717 (\AA) \\
    Free spectral range (km~s$^{-1}$) 	&	964\\
    Exposure time calibration 			& 0.5 s/channel \\
    Exposure time object 				&  120 s/channel\\
  \hline  
\end{tabular}
\end{table}

\section{Observations and data reductions}
Observations were made by M. Rosado with the f/7.5 Cassegrain focus of the 2.1~m telescope from the Observatorio Astron\'omico Nacional of the Universidad Nacional Aut\'onoma de M\'exico (OAN, UNAM), at San Pedro M\'artir, B. C., M\'exico on July 2015 using the UNAM scanning Fabry-Perot interferometer PUMA \citep{Rosado1995}. We used a 2048 $\times$ 2048 pixels Marconi CCD detector with a binning factor of 4, obtaining a $512\times512$ pixel window with a spatial sampling of 1''.3 pixel$^{-1}$. The Fabry-Perot sampling spectral resolution is 0.41 $\text{\AA}$ at H${\alpha}$ (equivalent to 19.0~km~s$^{-1}$) and a free spectral range of 19.8 $\text{\AA}$ (equivalent to a velocity range of 908~km~s$^{-1}$).

\begin{figure*}
 \includegraphics[width=1.7\columnwidth]{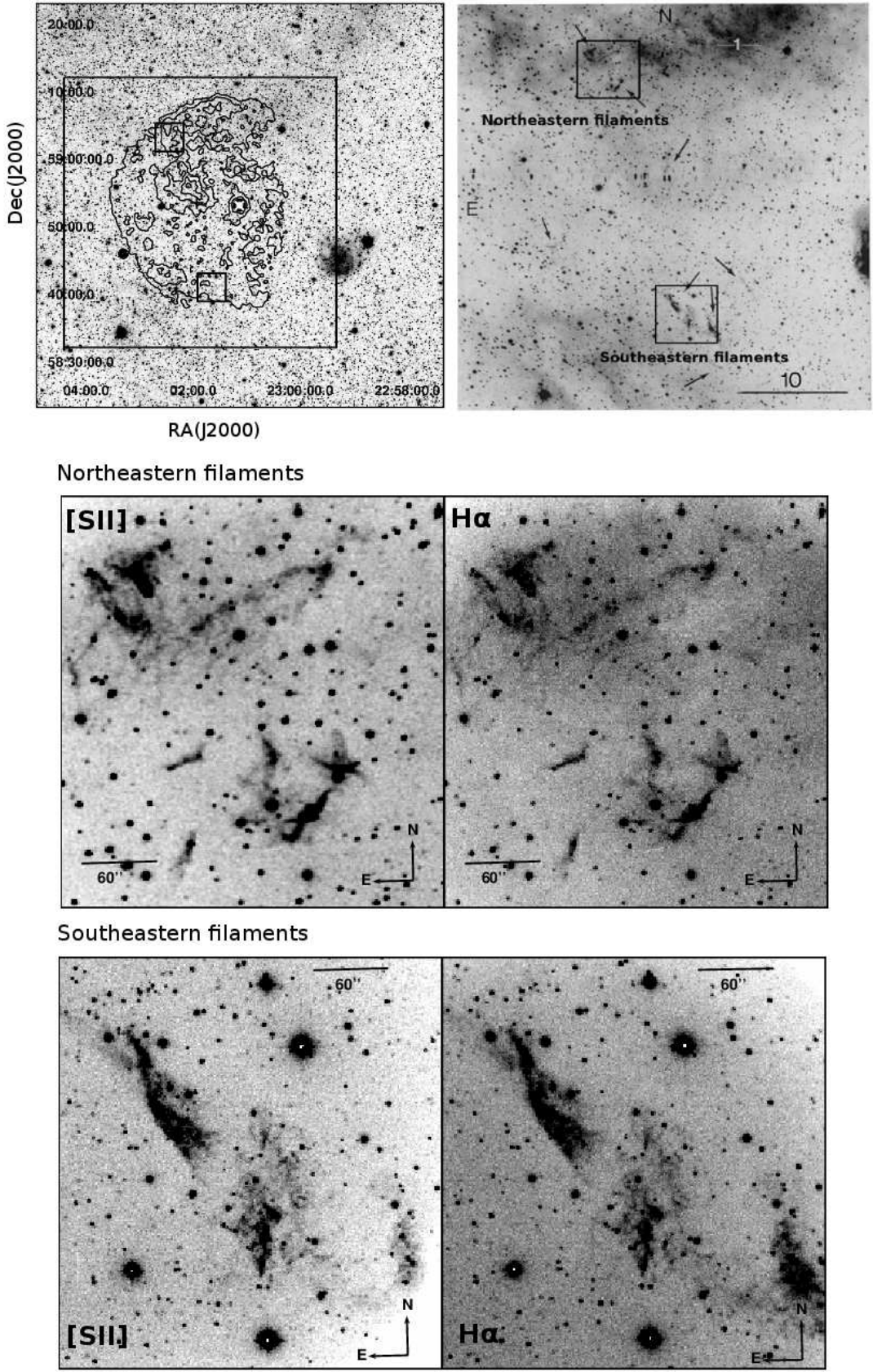}
  \caption{\textit{Top-left}: DSS image of the SNR CTB\,109. Overplotted is the X-ray emission from ROSAT \citep{Hurford-Fesen1995}. The big box indicates the size of the H$\alpha$ image of \citep{Fesen-Hurford1995}, the small boxes show the locations and the field of view of the regions studied in this work. \textit{Top-right}: The H$\alpha$ image published by \citep{Fesen-Hurford1995}. Squares represent the field of view of the regions studied in this work. \textit{Middle} and \textit{Bottom panels}: H$\alpha$ (right) and [SII] (left) direct images obtained with PUMA of northeastern and southeastern filaments (filaments A to H in Figure \ref{F3}).}
  \label{F2}
\end{figure*}

We obtained a set of direct images in H${\alpha}$ and [SII]$\lambda \lambda$6717,6731$\text{\AA}$ using the PUMA focal reducer without the Fabry-Perot interferometer in the instrument optical axis. The exposure time of each of the direct images was 120 s. 
We also obtained [SII]$\lambda \lambda$6717,6731 $\text{\AA}$ Fabry-Perot data cubes of two regions of the SNR CTB\,109 (see Figure~\ref{F2}) located to the north and south of the remnant and containing the filaments 1, 2 and 4 classified by \cite{Fesen-Hurford1995}. We scan the FP interferometer through 48 channels, with integration time of 120~s per channel obtaining an object data cube of dimensions $512\times512\times48$. The interference filter used was centred at 6720 $\text{\AA}$ with a bandwidth of 20 $\text{\AA}$ so that both [SII] lines at 6717 $\text{\AA}$ and 6731 $\text{\AA}$ are detected separated by 15 channels. We were able to compute pixel per pixel the [SII]$\lambda$6717/[SII]$\lambda$6731 line-ratio from the data cube. We also obtained a calibration data cube of dimensions $512\times512\times48$ reproducing the same observing conditions of the object data cube by introducing a mirror in the optical path of the interferometer. For calibrating the data cube we used a neon lamp (6717.04 $\text{\AA}$ wavelength calibration).  Observational and instrumental parameters are listed in Table~\ref{T2}.

The data reduction and analysis of the Fabry-Perot data cubes were performed using the software CIGALE \citep{LeCoarer1993} and ADHOC software. The direct images were reduced using IRAF\footnote[2 ] {IRAF is distributed by National Optical Astronomy Observatory, operated by the Association of 
Universities for Research in Astronomic, Inc., under cooperative agreement with 
the National Science Foundation.} tasks. The radial velocity profiles were fitted with a minimum number of Gaussian functions convolved with the instrumental function (Airy function).  The CIGALE data reduction process allows us to compute the parabolic phase map from the calibration cube. This map provides the reference wavelength for the line profile observed for each pixel in order to derive the wavelength map. From the phase map, the monochromatic, continuum and velocity maps were also computed. The process to extract the kinematic information is  described in \cite{Valdez2001,Sanchez2015}.

\section{Results}
\subsection{Direct images: H$\alpha$ and [SII] optical counterparts}
Figure \ref{F2} shows: 1) DSS image showing the  field of view of the SNR CTB\,109 with overplotted X-ray contours from ROSAT, 2) the H$\alpha$ image of \cite{Fesen-Hurford1995} showing the northeastern and southeastern filaments studied in this work and 3) our H$\alpha$ and [SII] direct images of those filaments, both show high [SII]/H$\alpha$ line-ratios.

\begin{figure*}
\includegraphics[width=2\columnwidth]{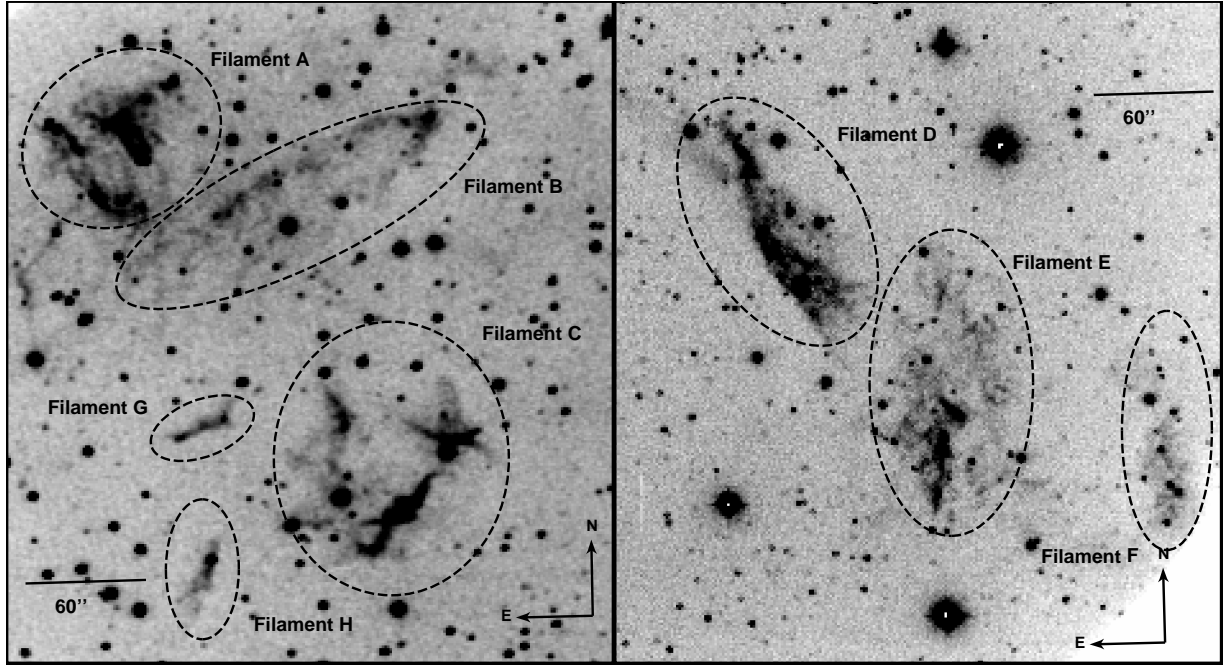}
  \caption{[SII]$\lambda \lambda$6717,6731$\text{\AA}$ direct images of the different filaments of the SNR CTB\,109 taken with PUMA in its direct image mode. Left panel corresponds to the northeastern filaments, while right panel corresponds to southeastern filaments.}
  \label{F3}
\end{figure*}

\begin{figure*}
\includegraphics[width=2\columnwidth]{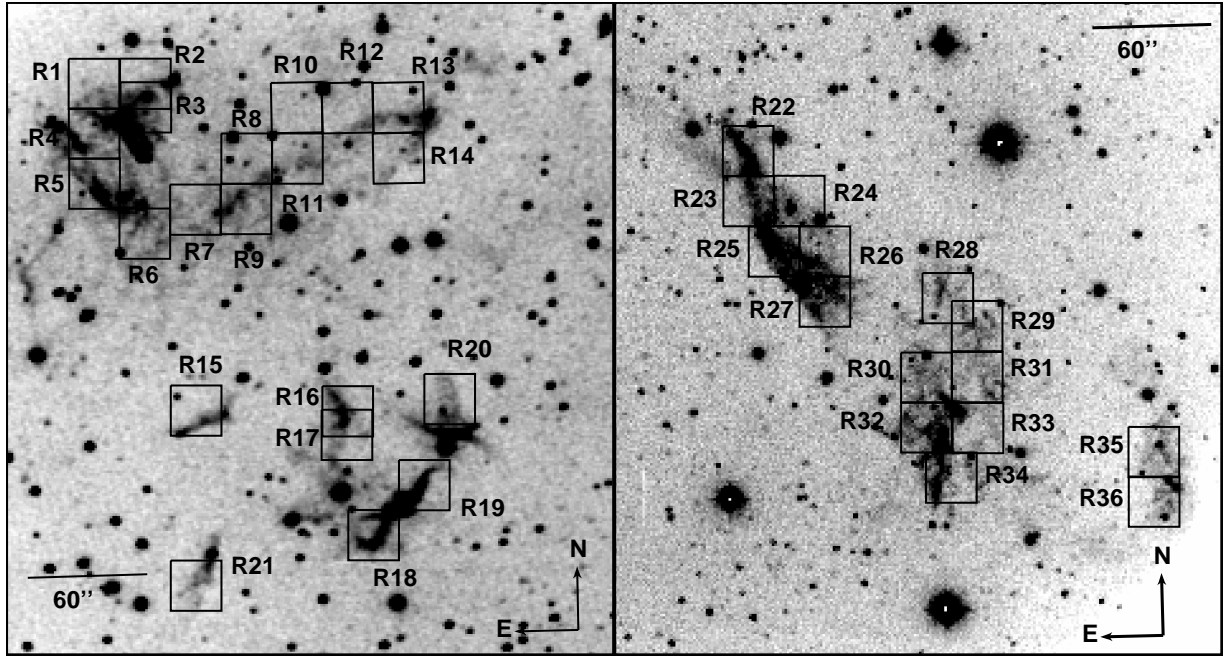}
  \caption{[SII]$\lambda \lambda$6717,6731$\text{\AA}$ direct images of the filaments of the SNR CTB\,109 showing the regions in which we divided the filaments in order to carry out our analysis (R1 to R36). Left and right panels correspond to the northeastern and  southeastern filaments, respectively. }
  \label{F4}
 \end{figure*}

\begin{figure*}
\includegraphics[width=2\columnwidth]{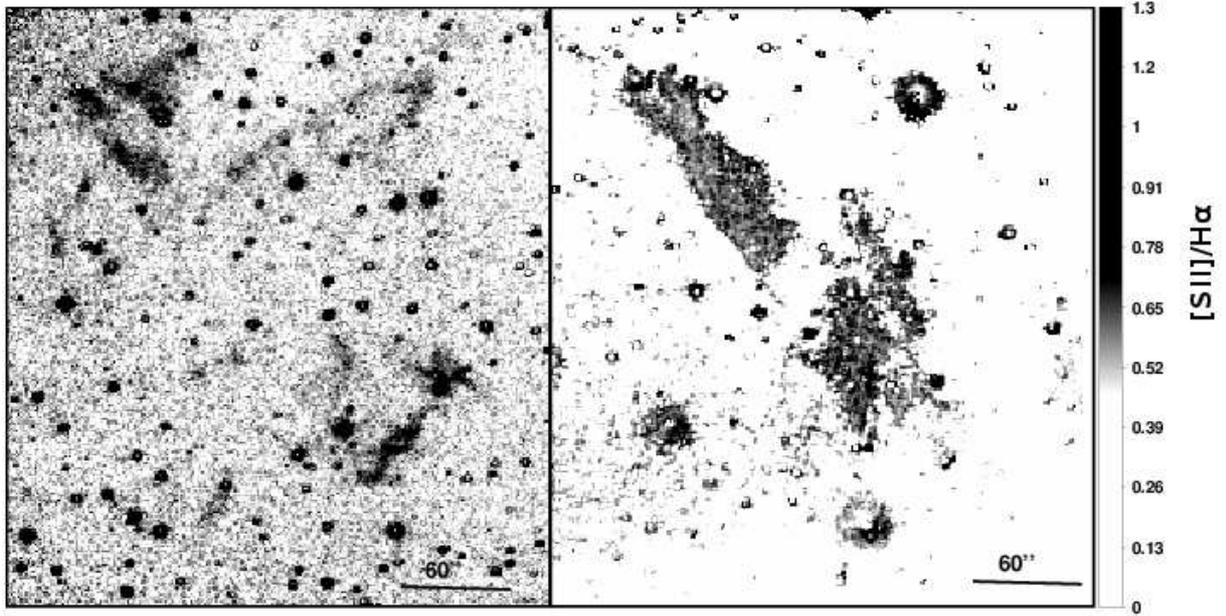}
  \caption{[SII]/H$\alpha$ map of northeastern (left) and southeastern (right) filaments of the SNR CTB\,109. It is shown that the [SII]/H$\alpha$ line ratio value are larger than 0.5, characteristic of SNR.}
  \label{F5}
\end{figure*}

Figure \ref{F3} shows the [SII] direct images with our filament labels (Filament A to Filament H). Filaments A, B and D shown in Figure \ref{F3} were previously classified as filament 1, 2 and 4, respectively by \cite{Fesen-Hurford1995}. Figure~\ref{F4} shows the 20$\times$20 pixels (6.6''$\times$6.6'') regions used for our analysis (regions R1 to R36).  The image of [SII]/H$\alpha$ 2D line-ratios is depicted in Figure \ref{F5}. The [SII]/H$\alpha$ line ratios are mostly above 0.6 showing that the emission is coming from the shock heated gas indicative of  supernova remnant emission. From this figure we can see that the [SII]/H$\alpha$  line-ratio values of the filaments span from 0.61 to 1.37 for filaments A, B, C and D (see Table 3 for more detailed values).
For filaments E, F, G and H the [SII]/H$\alpha$ line-ratios are less than one but still larger than 0.6. There are other non filamentary regions with [SII]/H$\alpha$ ratios of about 0.4. The values of the northeastern filaments agree with those obtained by \cite{Fesen-Hurford1995} (1.12 and 1.07 for their filaments 1 and 2 which correspond to filaments A and B in this work). For filament D, the value is similar (0.82) to the value of their filament 4. 

From the spectroscopic study of \cite{Fesen-Hurford1995} and from our measured [SII]/H$\alpha$ 2D line ratios we demonstrate that these filaments are indeed the optical counterpart of the SNR CTB\,109.

\begin{figure*}
\includegraphics[width=2\columnwidth]{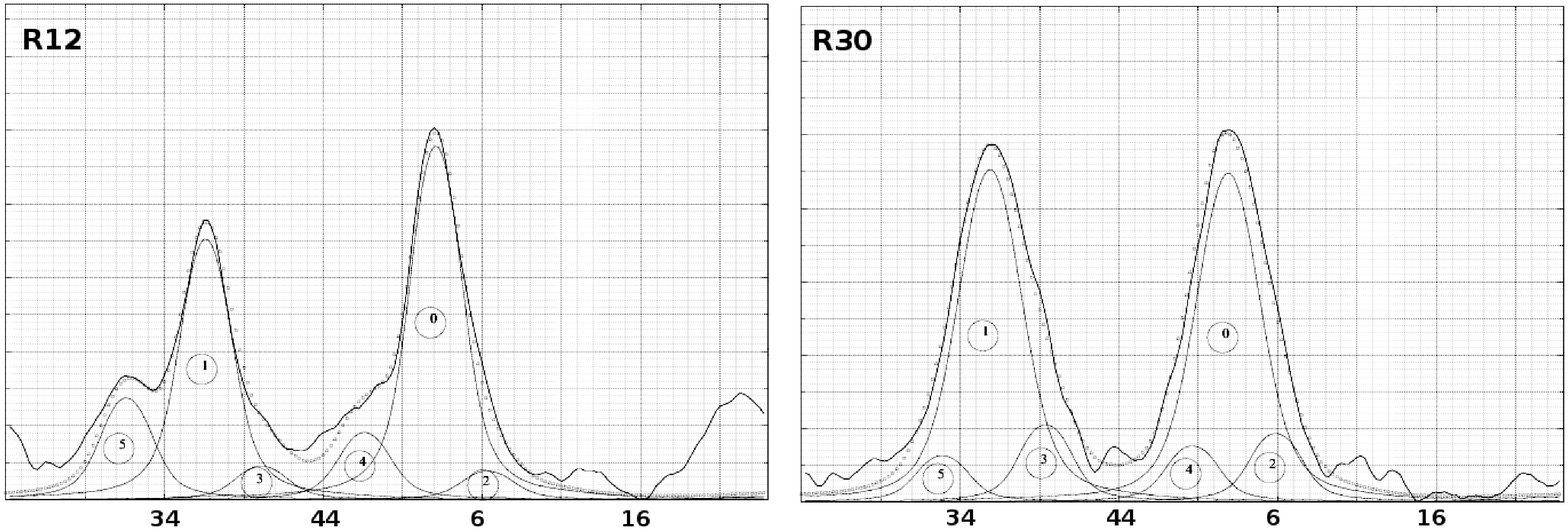}
 \caption{[SII] line profiles of regions 12 and 30 obtained with Fabry-Perot. The profiles were integrated  over boxes with a 20$\times$20~pixel size.  The x-axis in the line profiles gives the lambda channels and the y-axis is the intensity in arbitrary units. Both [SII] lines at 6717 $\text{\AA}$ and 6731 $\text{\AA}$ are detected. Decomposition of each profile is indicated in thin lines. Resulting profile is shown as hollow circles and with numbers. Dotted line represent the sum of all fitted components.}
  \label{F6}
\end{figure*}

\subsection{Kinematics of the SNR CTB\,109}

\subsubsection{Systemic Velocity and Kinematic Distance}

From the scanning FP interferometric data cubes in [SII]$\lambda \lambda$6717,6731~$\text{\AA}$ we obtained the line profiles of regions R1 to R36. Each profile was integrated over a 20$\times$20~pixels square. Figure~\ref{F6} shows the [SII]$\lambda \lambda$6717,6731~$\text{\AA}$ line profiles of regions \#12 and \#30 as an example. The [SII] lines, $\lambda$6717~$\text{\AA}$ and $\lambda$6731~$\text{\AA}$ are detected. They lie about 15 velocity channels apart (equivalent to about 285 km~s$^{-1}$). The $\lambda$6731 $\text{\AA}$ line is fainter in most of the cases. Since both lines  come from the same gas layer, we will focus on the brighter line ([SII]$\lambda$6717 $\text{\AA}$) that has higher S/N-ratio. From Figure \ref{F6} we can see that both [SII] lines present more than one velocity component. Each of these profiles can be fitted with a Gaussian whose maxima is associated with the velocity along the line of sight of that parcel of gas.

We assume the brightest component of the line to be the main velocity component of the SNR. It is marked with No.~0 in Figure \ref{F6}. For each emission line we can also identify two extreme velocity components. We shall refer to these as V$_{min}$ and V$_{max}$. These are marked with No.~2 and No.~4, respectively. 

Table~\ref{T3} reports the values for the velocity components of the [SII]$\lambda$6717~$\text{\AA}$ line. Column 1 lists the region ID. Columns 2, 3 and 4 indicate the main and extreme velocity components (V$_{min}$ and V$_{max}$) corrected for the Local Standard of Rest (LSR). Column 5 gives the difference in velocity between the extreme velocity components. Column 6 lists the distance of each region in arcmin to the geometrical centre of the SNR, Column 7 lists the [SII]/H$\alpha$ line-ratio and finally Column 8 indicates the associated filament to which each region belongs to.

As one can see from Figure \ref{F6} and Table~\ref{T3}, the radial velocity profiles are, in general, composite implying large motions as expected in a SNR. They show at least three velocity components: The brightest one corresponds to the systemic motion of the SNR, while two faint velocity components are seen at extreme velocities. 
The several velocity components found in the velocity profiles of the filaments are probably due to the supernova remnant shock expansion, as well as to contamination due to the interaction of the SNR gas with dense interstellar clouds that could be associated with the HII regions located to the north and southwest of the remnant. 
It is interesting to note the existence in the SNR velocity profiles of a velocity component also identified with those of regions located in the Perseus arm, as we will discuss below.

The velocity values of the main component vary from -34~km~s$^{-1}$ to -89~km~s$^{-1}$. Taking the average of all the velocity values of the main component in Table 3 we obtain a LSR velocity of  -59$\pm$17~km~s$^{-1}$  as a first estimate of the systemic velocity of the SNR. The average values for the extreme velocity components, in all the boxes where velocity profiles were extracted, are: V$_{min}$ $\sim$-122~km~s$^{-1}$  and  V$_{max}$ $\sim$+152~km~s$^{-1}$. Another way of obtaining the systemic velocity is to average the extreme velocity components (V$_{min}$ and V$_{max}$). Doing this we obtained a systemic velocity V$_{LSR}$=-50$\pm$6~km~s$^{-1}$. For some regions we find an additional velocity component averaging 1~km~s$^{-1}$. This later velocity component agrees with the component found by \cite{Roberts1972}  in the theorical HI profile of the TASS model at 12$^{\circ}$ pitch angle (V$_{LSR}$=0~km~s$^{-1}$) in the direction of the SNR CTB\,109.  \cite{Roberts1972} attributed this component to gas located  very close to the Sun in the ``Orion spur''.

Figure \ref{F7} shows a velocity-position diagram of the several velocity components of the SNR CTB\,109. We note that the extreme velocities define the envelope of the expansion velocity ellipse (marked with ``x'') which will be discussed in Section 3.2.3. The dashed line depicts the value of the systemic velocity V$_{LSR}$=-50$\pm$6~km~s$^{-1}$ we found by averaging the extreme velocity component values.

The systemic velocity found for the SNR CTB\,109 from its optical filaments is similar to the velocity ranges found for HII regions in the Perseus arm and to the velocity values of neutral and molecular material found in the direction of the SNR and in neighbouring molecular clouds.

\subsubsection{Distance estimate}

As we have seen in the Introduction the distance determination to the  SNR CTB\,109 has been done by taking into account different methods. Also, the kinematic ambiguity has been solved by taking into account different possible scenarios based in related emission of the surrounding medium of the supernova remnant and its central magnetar. 

In this work we used the radial velocity of the optical filaments of the  SNR CTB\,109 to determine the distance to the supernova remnant by using the velocity-distance diagram of \cite{Kothes-Foster2012}.  In this diagram, given our radial velocity (-50~km~s$^{-1}$) there are three possible values of the distance: The first one located in the shock jump at 3 kpc, the second one located at the  downstream shock (d=3.2 kpc) and the last one located in the interarm region (d=4 kpc). To solve this ambiguity, we considered the  distance  to the magnetar AXP~1E~2259+586 determined using the dispersion measure (DM). The distance to the magnetar AXP~1E~2259+586 was obtained  by using the \cite{Cordes-Lazio2002} electron density model (NE2001 model). Considering  the frequency range of 11.24-110.6 MHz and dispersion measure of DM=79$\pm$4 pc cm$^{-3}$ given by \cite{Malofeev2006}, we found that the distance to the magnetar is about 2.8$\pm$0.1 kpc. Therefore, the most likely distance to the supernova remnant is 3 kpc. The 4~kpc distance estimate has been already eliminated by \cite{Kothes-Foster2012} arguments. The d=3.2 kpc value favoured by \cite{Kothes-Foster2012} is more difficult to eliminate because, as pointed out by those authors, the geometrical centre of the SNR does not coincide with the magnetar position implicating a migration of the magnetar in a time of about 10$^{4}$~yr.  
Our systemic velocity determination together with the DM constraint favour a distance of 3 kpc for the SNR CTB\,109 locating it in the Perseus arm. The possibility that the distance is slightly larger (3.2 kpc) cannot be ruled out. In any case the SNR is located in an spiral arm and its progenitor is a massive star. Thus, we will adopt a distance of 3.1$\pm$0.2 kpc considering this distance indetermination due to kinematics and taking into account the different aspects already mentioned including the migration of the progenitor downstream.

Our distance determination to CTB\,109 is based considering that the Perseus arm is located at 3 kpc. Nevertheless, the current electron density model of \cite{Yao2017} (YMW16 model)  reveals that the distance to the magnetar AXP~1E~2259+586 using the dispersion measure DM=79$\pm$4 pc cm$^{-3}$ given by \cite{Malofeev2006} is about 2.46$\pm$0.09 kpc. This value suggests the possibility that the Perseus arm is closer to what has been thought as proposed previously by \cite{Roberts1972,Xu2006}.

\begin{table*}\centering
\begin{minipage}{85mm}
	\caption{Velocities of Filaments of the SNR CTB\,109}
	\label{T3}
\begin{tabular}{lcccccccc}
\hline
 Region &
 V$_{LSR}$ & 
 V$_{minLSR}$ & 
 V$_{maxLSR}$ & 
 $\Delta V$ &
 r (arcmin) &
[SII]/H$\alpha$ &
 Filament\\
 (20$\times$20 pix) &
 km~s$^{-1}$ & 
 km~s$^{-1}$ &
 km~s$^{-1}$ &
 km~s$^{-1}$ &
 &
 &location\\
\hline
R1	&	-46	&	29	&	-109	&	138	&	14.50	&	1.31	&	FIL A	\\
R2	&	-32	&	39	&	-96	&	135	&	14.26	&	0.98	&	FIL A	\\
R3	&	-43	&	36	&	-127	&	163	&	14.09	&	0.98	&	FIL A	\\
R4	&	-19	&	59	&	-108	&	167	&	14.15	&	1.31	&	FIL A	\\
R5	&	-16	&	87	&	-76	&	163	&	13.80	&	1.06	&	FIL A	\\
R6	&	-58	&	103	&	-130	&	233	&	13.21	&	1.14	&	FIL A,B	\\
R7	&	-64	&	116	&	-156	&	272	&	13.14	&	1.09	&	FIL B	\\
R8	&	-62	&	-1	&	-150	&	149	&	13.27	&	1.14	&	FIL B	\\
R9	&	-64	&	2	&	-147	&	149	&	12.91	&	1.14	&	FIL B	\\
R10	&	-62	&	6	&	---	&	---	&	13.44	&	1.30	&	FIL B	\\
R11	&	-62	&	8	&	-158	&	166	&	13.06	&	0.78	&	FIL B	\\
R12	&	-64	&	1	&	-150	&	151	&	13.24	&	0.70	&	FIL B	\\
R13	&	-66	&	-3	&	-141	&	138	&	13.06	&	0.88	&	FIL B	\\
R14	&	-70	&	-12	&	-139	&	127	&	12.67	&	0.70	&	FIL B	\\
R15	&	-49	&	60	&	---	&	---	&	11.74	&	0.61	&	FIL G	\\
R16	&	-53	&	96	&	---	&	---	&	11.00	&	0.99	&	FIL C	\\
R17	&	-54	&	122	&	---	&	---	&	10.84	&	0.36	&	FIL C	\\
R18	&	-52	&	152	&	-119	&	271	&	10.01 &	0.40	&	FIL C	\\
R19	&	-44	&	34	&	-113	&	147	&	10.13	&	0.91	&	FIL C	\\
R20	&	-63	&	41	&	---	&	---	&	10.65	&	1.37	&	FIL C	\\
R21	&	-38	&	51	&	-156	&	207	&	10.61	&	0.40	&	FIL H	\\
R22	&	-79	&	-27	&	-119	&	92	&	10.90	&	0.88	&	FIL D	\\
R23	&	-82	&	-9	&	-131	&	122	&	11.32	&	1.02	&	FIL D	\\
R24	&	-74	&	-13	&	-132	&	119	&	11.27	&	0.72	&	FIL D	\\
R25	&	-89	&	-32	&	-129	&	97	&	11.72	&	0.75	&	FIL D	\\
R26	&	-87	&	-26	&	-143	&	117	&	11.69	&	0.56	&	FIL D	\\
R27	&	-83	&	-11	&	-140	&	129	&	12.11	&	0.93	&	FIL D	\\
R28	&	-57	&	-10	&	-121	&	111	&	12.07	&	0.81	&	FIL E	\\
R29	&	-71	&	-14	&	-118	&	104	&	12.32	&	0.80	&	FIL E	\\
R30	&	-68	&	-10	&	-113	&	103	&	12.75	&	0.61	&	FIL E	\\
R31	&	-53	&	48	&	-136	&	184	&	12.75	&	0.44	&	FIL E	\\
R32	&	-62	&	19	&	-138	&	157	&	13.16	&	0.67	&	FIL E	\\
R33	&	-64	&	-4	&	-120	&	116	&	13.18	&	1.06	&	FIL E	\\
R34	&	-61	&	14	&	-124	&	138	&	13.59	&	1.08	&	FIL E	\\
R35	&	-45	&	11	&	-96	&	107	&	13.54	&	0.56	&	FIL F	\\
R36	&	-42	&	25	&	-102	&	127	&	13.95	&	0.60	&	FIL F	\\
\hline
\end{tabular}
\end{minipage}
\end{table*}

\subsubsection{Expansion Velocity}

\begin{figure*}
\includegraphics[width=1.8\columnwidth]{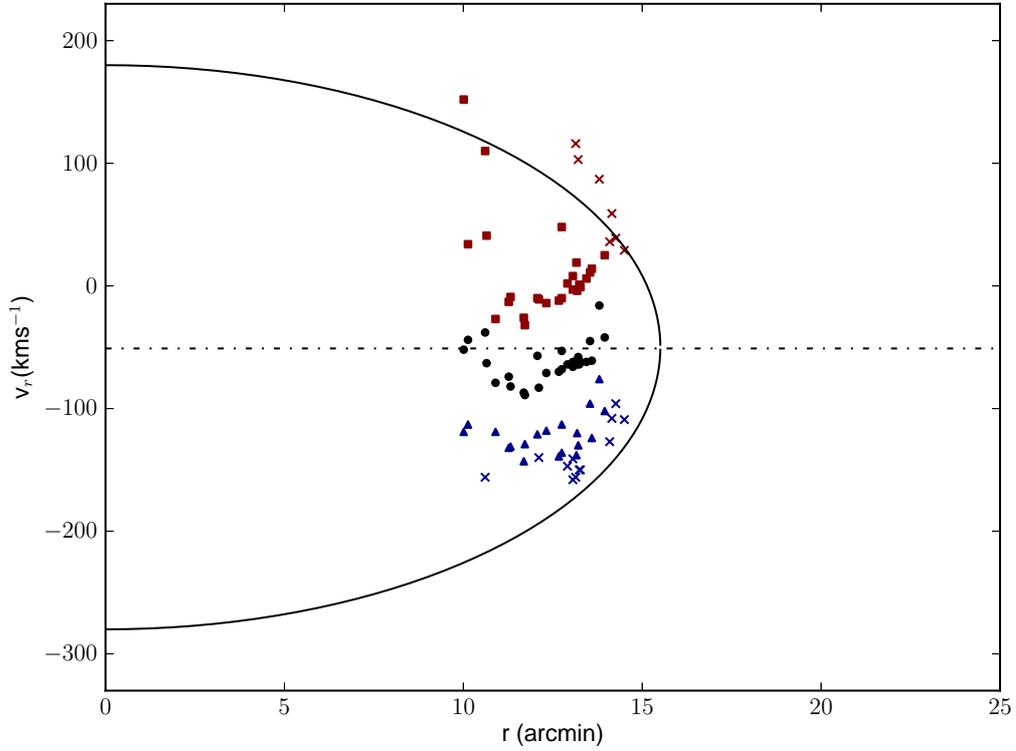}
  \caption{Position-velocity diagram of the northeastern and southeastern filaments. The X-axis corresponds to the distance (in arcmin) to the remnant centre obtained using Figure \ref{F2}. The continuum line is the Doppler ellipse. The dot marks (in black) represent the main velocity component of the fitted profile, the extreme velocity component V$_{max}$ and V$_{min}$ are in in red and blue, respectively. The ``x'' marks represent the regions that define the envelope of the velocity ellipse. Dotted line represents the adopted  systemic radial velocity of the supernova remnant (-50~km~s$^{-1}$).}
 \label{F7}
\end{figure*}

In order to determine the expansion velocity of the SNR we plotted the velocity components versus position listed in Table~\ref{T3}.  Since we are considering that the remnant is a thin shell expanding radially with a constant expansion velocity  V$_{exp}$, the projection of the plane gives us an ellipse called the ``Doppler ellipse''. This ellipse is shown in Figure \ref{F7}. As mentioned before, from this figure we can see that the points fall close to the extreme of the Doppler ellipse that corresponds to the largest radii. In this Figure other velocity components can also be identified: The systemic velocity (circles), the 0 LSR velocity of the Orion spur (squares), and finally the extreme velocities that define the velocity ellipse. The regions that define the envelope of the velocity ellipse are marked with ``x'' and correspond to regions R1, R2, R3, R4, R5, R6, R7, R8, R9, R11, R12, R13, R21 and R27. Note that these regions have in general [SII]/H$\alpha$ ratios larger than one. To determine the expansion velocity we considered the velocity values given by the intersection of the Doppler ellipse with the y-axis. We find an expansion velocity for the SNR of about 230$\pm$5~km~s$^{-1}$.

 \begin{figure*}
\includegraphics[width=2\columnwidth]{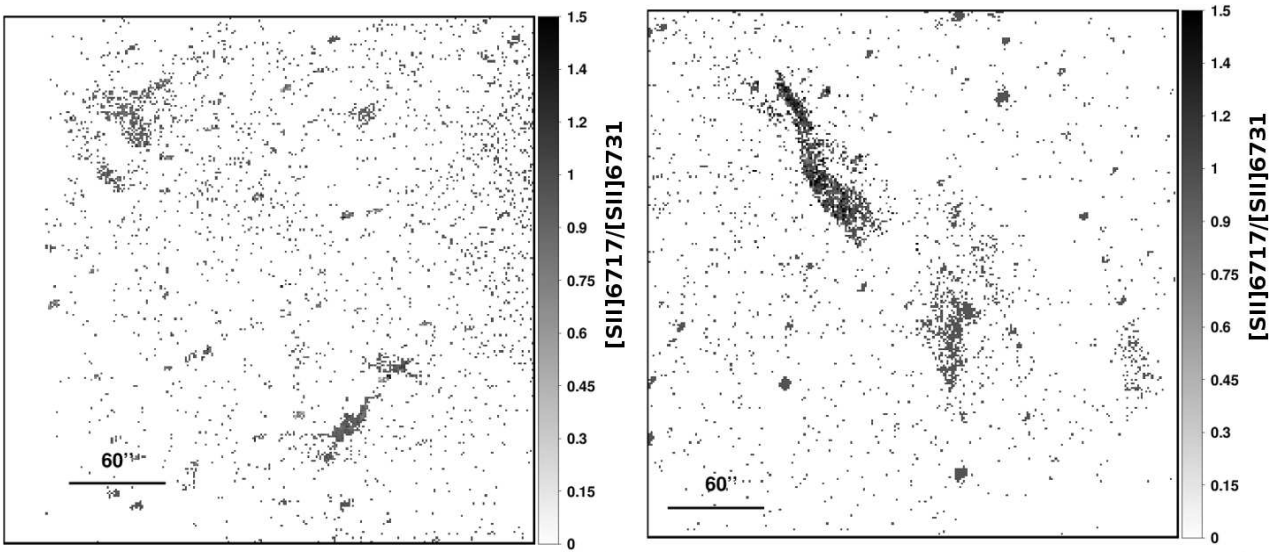}
  \caption{[SII]$\lambda$6717/[SII]$\lambda$6731 ratio images of filaments A to H (see Figure \ref{F3}) of the SNR CTB\,109.}
 \label{F8}
\end{figure*}

%\hspace{3cm}

\subsubsection{Electron density}
We used the [SII]$\lambda$6717/[SII]$\lambda$6731 line-ratios to determine the electronic density (n$_e$). The [SII]$\lambda$6717~$\text{\AA}$ and [SII]$\lambda$6731~$\text{\AA}$ lines are emitted by a single ion from two levels with the same energy but with different transition probabilities.
This optical line ratio is sensitive
to changes in the electron density and weakly dependent on the electron temperature. It is important to note that, despite the fact that X-ray emission implies that there is gas at very high temperature in the SNR CTB\,109 (of millions of K), the optical emission of this SNR should come from ionized gas with colder temperatures (about thousands of K). The line-ratios and velocities measured in this work show that this optical emission is indeed the optical counterpart of the X-ray emitting SNR. We can reconcile this difference in temperatures  by proposing that the primary shock wave of the SNR CTB\,109 is moving through a low density medium where dense cloudlets are immersed inducing a secondary shock wave in the dense cloudlets, as proposed by \cite{McKee-Cowie1975}. The primary shock wave emits the X-rays while the induced shock wave emits at optical wavelengths. The velocities of the different shocks (primary and induced) can be related using the following relations:

\begin{equation}
\label{eq1}
n_{opt} V_{opt}^{2}\sim n_x V_x^{2} \ \ \ \ \ \ \ \ Sedov-Taylor
\end{equation}
\begin{equation}
\label{eq2}
2 V_{opt}\sim V_{x'}\ \ \ \ \ \ \ \ \ \ \ Radiative 
\end{equation}

where V$_{x}$  is the velocity of the primary blast wave in the Sedov-Taylor phase, and V$_{x'}$  is the velocity of the primary blast wave in the radiative phase.
This model also explains why we can have a SNR in its Sedov-Taylor or radiative phase while it also emitting in optical wavelengths.
$n_{opt}$ is the pre-shock electron density, $n_x$ is the density of the low density medium and  V$_{opt}$ is the of the induced blast wave.
 
The [SII]6717/[SII]6731 line-ratio values of the filaments are: 1.1 for filament A, 0.8 for filaments B and C, 0.9 for filament D, 1.2 and 1.8 for filaments E and F, respectively and for filaments  G and H the values are 0.6 and 0.4, respectively. The values of filaments A and B are the same that those obtained by \cite{Fesen-Hurford1995}. On the other hand, the [SII]6717/[SII]6731 line-ratio value of filament D is lower than their value. 

[SII]$\lambda$6717/[SII]$\lambda$6731 ratio images are shown in Figure~\ref{F8}. These images were done using ADHOCw\footnote[3]  {developed by J. Boulesteix} software to extract from the lambda datacube two subdata cubes corresponding to the [SII]$\lambda$6717 $\text{\AA}$ and [SII]$\lambda$6731 $\text{\AA}$ lines.  The sum of all the channels from each subdata cube results in two 2D images one of each [SII] line. 

To determine n$_e$, we assumed that the S$^{+}$ emission is formed in a region between 5000 K and 10$^{4}$~K (which corresponds to typical temperature range of optical emission nebulae). Using  the \textit{temden} routine of STSDAS/IRAF and the average of the [SII]$\lambda$6717/[SII]$\lambda$6731 ratio, we found a n$_e$ for both the northeastern and southeastern filaments of n$_e$=501$\pm$120~cm$^{-3}$ (for T$_e$=5$\times$10$^{3}$~K) and n$_e$=658$\pm$250~cm$^{-3}$ (for T$_e$=10$^{4}$~K), respectively. The average of these values is 580$\pm$185~cm$^{-3}$.

Our electron density values at T=5000-10000~K are within the range of values n$_e$=220-800~cm$^{-3}$ presented by \cite{Fesen-Hurford1995}.

If we consider that the shock is radiative in the cloudlet (where the compressed gas is cooled by collisions), the pre-shock electron density n$_{opt}$ is given by:

\begin{equation}
\label{eq3}
n_{opt} =n_e\left(\frac{c_s}{V_s}\right)^2
\end{equation}

where c$_s$=10~km~s$^{-1}$ is the sound speed of the environment at T$_e$ (in this work we use T$_e$=5000 K and T$_e$=10$^{4}$~K), V$_s$=$V_{exp}$=230$\pm$5~km~s$^{-1}$. Taking n$_e$=580$\pm$185~cm$^{-3}$, we found a pre-shock density n$_{opt}\approx$1.1$\pm$0.3~cm$^{-3}$.

If we consider that the low density medium where the primary blast wave evolves has a density n$_x$=0.16~cm$^{-3}$ \citep{Sasaki2004}, then equations \ref{eq1} and \ref{eq2} imply that V$_x$=603~km~s$^{-1}$ (for Sedov-Taylor phase) and V$_{x'}$=460~km~s$^{-1}$ (for radiative phase), and the X-ray emission is explained.

\subsubsection{The Energy and Age of the SNR CTB\,109}

We can determine the age and energy of the SNR either if the SNR is in the Sedov-Taylor or in its radiative phase of evolution following \cite{McKee-Cowie1975} ideas discussed before for explaining the optical emission. 

The age of the SNR can be obtained with the  numerical models of  \cite{Cox1972} (for the Sedov-Taylor phase) or  \cite{Chevalier1974} (for the radiative phase):

\begin{eqnarray}
\label{eq4}
t(4)=\left\{
\begin{array}{l@{\quad}l}
    39.15 R/V_{x}   \ \ \ \ \ \ \ \ Sedov-Taylor\\
    \\
   30.7 R/V_{x'}   \ \ \ \ \ \ \ \ \ \ \ Radiative 
   \end{array}
   \right.
\end{eqnarray}

where t(4) is the age of the remnant in units of 10$^{4}$~yr, V$_{x}$ and V$_{x'}$ are the velocities of the primary shock wave according to the \cite{McKee-Cowie1975} scenario in the Sedov-Taylor and in the radiative phases, respectively. Since the shock velocity induced in the cloudlets is $V_{exp}$=V$_{opt}$=230~km~s$^{-1}$, as we have determined in this work and using the relation: n$_{opt}$V$_{opt}^{2}\sim$ n$_x$V$_x^{2}$, giving the primary blast wave velocity: V$_x$=603~km~s$^{-1}$ and V$_{x'}$=460~km~s$^{-1}$, R is the linear radius, 13.9 pc,  adopted from the radius of the circular X-ray emission considering the distance of 3.1 kpc that we have determined in this work.  

Thus, the age obtained in this work for the  Sedov-Taylor phase is 9.0$\times$10$^{3}$~yr and for the radiative radiative phase is 9.2$\times$10$^{3}$~yr.

Those values are in agreement with the values obtained by several works studying the X-ray emission of the SNR as listed in Table \ref{T1}

The energy deposited in the ISM by the SN explosion can be determined with the numerical models of  \cite{Cox1972} (Sedov-Taylor phase) or  \cite{Chevalier1974} (for radiative phase):

\begin{equation}
\label{eq5}
E_{51}=2.44\times10^{-9}n_x^{1.0}V_x^{2}R^{3} \ \ \ \ \ \ \ \ \ \ \ \ Sedov-Taylor
\end{equation}

\begin{equation}
\label{eq6}
E_{50}=5.3\times10^{-7}n_x^{1.12}V_{x'}^{1.4}R^{3.12}  \ \ \ \ \ \ \ \ \ \ \ \ Radiative 
\end{equation}

where E$_{50}$ and  E$_{51}$ are in units of 10$^{50}$~erg and 10$^{51}$~erg, respectively. V$_{x}$  and V$_{x'}$ are the shock velocities in km~s$^{-1}$ in the low density medium where the primary blast wave evolves, in the Sedov-Taylor and in the radiative phases, respectively. n$_x$ is the pre-shock electron density in cm$^{-3}$,  R is in pc. For the Sedov-Taylor phase one can take n$_x^{1.0}$V$_x^{2}$ from our optical observations directly because the McKee \& Cowie scenario shows that n$_{opt}$V$_{opt}^{2}\sim$ n$_x$V$_x^{2}$.

The value of the energy deposited in the ISM by the SN explosion derived here is E$_0$=5.2$\times$10$^{50}$ erg if the SNR is in its Sedov-Taylor phase and E$_0$=1.8$\times$10$^{50}$ erg if the SNR is in its radiative phase of evolution. Table~\ref{T4} shows the kinematical parameters determined for the SNR CTB\,109.
 
The energy values obtained for the SNR either in the Sedov-Taylor phase or in its radiative phase are lower than the value obtained by \cite{Rho-Petre1997} and similar to  that obtained by \cite{Sasaki2004}. In any case, the energy values derived here are typical of supernova producing typical pulsars, so that, the energy released in the supernovae explosion cannot be used to identify whether a magnetar is produced.

\begin{table}
	\centering
	\caption{Kinematic Parameters derived for the SNR CTB\,109 in this work.}
 \label{T4} 
 \begin{tabular}{llcc}
 \hline
Parameter & 
Value\\
 \hline
 V$_{sys}$ (LSR)	& -50$\pm$6 km~s$^{-1}$\\
 V$_{exp}$			& 230$\pm$5 km~s$^{-1}$\\
 n$_{opt}$ 			& 1.1 cm$^{-3}$ \\
 V$_x^{a}$			& 603 km~s$^{-1}$\\
 V$_{x'}^{b}$		& 460 km~s$^{-1}$\\
 n$_x^{c}$			& 0.16 cm$^{-3}$ \\
\hline
 \multicolumn{2}{c}{Sedov-Taylor phase}\\
 \hline
 Age 		& 9.0$\times$10$^{3}$~yr &  \\
 E$_0$ 			& 5.2$\times$10$^{50}$ erg\\ 
\hline
\multicolumn{2}{c}{Radiative phase} \\
\hline
 Age 		& 9.2$\times$10$^{3}$~yr & \\
 E$_0$ 			& 1.8$\times$10$^{50}$ erg\\ 
 \hline
\end{tabular}\\
\textit{a}~V$_x$ is the primary shock wave velocity in Sedov-Talor phase considering the McKee \& Cowie scenario.\\
\textit{b}~V$_{x'}$ is the primary shock wave velocity in radiative phase considering the McKee \& Cowie scenario.\\
\textit{c}~n$_x$ taken from \citep{Sasaki2004}.\\
\end{table}

\subsubsection{Magnetar braking index based on the SNR age and its implications}
The braking index \textit{n} is a quantity related to a pulsar's rotational evolution \citep{Gao2016}. 
These authors determined the average braking indices for magnetars with ages of related SNRs with the expression  $n \approx 1+P/\dot{P}t_{SNR}$.
where \textit{P} is the  pulsar period, $\dot{P}$ is the derivative of \textit{P} and t$_{SNR}$ is the SNR age. 

The average braking index should be \textit{n=3} considering that only magneto-dipole radiation causes the pulsar spin-down. If winds are also in place \textit{n<3}, according with \cite{Gao2016}. A braking index \textit{n=5} suggests gravitational waves emission, while \textit{n$\gg$3} points to a magnetic field decay on the surface crust of the neutron star, according with those authors. Considering the \cite{Gao2016} braking indices determination for magnetars and taking our age determination of CTB\,109 (t=9.0$\times$10$^{3}$~yr), we obtain a braking index of about n=49-50. This value is similar to the braking index value of the pulsar PSR~B2148+52 \cite[n=49.6,][]{Alpar-Baykal2006}. Thus, for the magnetar inside CTB 109 (AXP~1E~2259+58) its  braking index points towards high magnetic field decay.

\section{Conclusions}

We have studied the $H\alpha$ and [SII]$\lambda \lambda$6717,6731~$\text{\AA}$ emission and the kinematics of the Galactic supernova remnant G109.1-1.0~(CTB\,109). The kinematic results obtained allow us to determine the distance to the remnant, its age and the energy deposited in the ISM by the supernova explosion. The primary results are summarized as follows:

1. We have shown that the optical filaments detected to the NE and SE are the optical counterparts of the SNR. We have labelled eight H$\alpha$ filaments (Filament A to Filament H), three of them classified previously by \cite{Fesen-Hurford1995}.  High [SII]/H$\alpha$ line-ratios confirmed that these H$\alpha$ filaments are the optical counterpart of the radio SNR. 

2. We also estimated the radial velocity components referred to the LSR for each filament. We found that the systemic velocity of the SNR is V$_{LSR}$=-50$\pm$6~km~s$^{-1}$  by averaging the extreme velocity values of the [SII] line profiles. 
This velocity value is similar to the velocities estimated for ionized, neutral and molecular gas in the direction of the SNR and probed to belong to the Perseus arm. It is important to remark however that this work reports for the first time velocities of filaments that really belong to the SNR due to their high [SII]/H$\alpha$ line-ratios and velocity broadening.

3. With this radial velocity we derived a distance to the SNR CTB\,109 of 3.1$\pm$0.2 kpc. This value is in agreement with that obtained by \cite{Kothes-Foster2012} and with the distance to the magnetar AXP~1E~2259+586 estimated using its dispersion measure (DM) of \cite{Malofeev2006} and the electron density model NE2001 \citep{Cordes-Lazio2002}. However, the current electron density model YMW16 \citep{Yao2017} places the magnetar closer giving the possibility that the Perseus arm is closer.

4. Considering the extreme radial velocity values (V$_{min}$ and V$_{max}$) of the  regions, we obtained an expansion velocity  V$_{exp}$=230$\pm$5~km~s$^{-1}$.

5. Using the [SII]$\lambda$6717/[SII]$\lambda$6731 ratios we found a mean electron density of n$_e$=580$\pm$185~cm$^{-3}$ (for a T=5000-10$^{4}$~K)
and a pre-shock density n$_{opt}\approx$1.1~cm$^{-3}$.

6. In order to explain the optical emission we considered the \cite{McKee-Cowie1975} model of a high velocity primary shock wave moving in a low density medium with dense clumps and inducing secondary shocks in the dense clumps, responsible of the optical emission while the X-ray emission comes from the low density inter-clump medium where the primary shock wave moves. Following this scenario it is possible to reconcile the optical emission found here with the possible evolutionary phases of the SNR CTB\,109: Sedov-Taylor or Radiative. 

7. Using the McKee \& Cowie model we computed the age of the SNR and the  energy deposited in the ISM by the SN explosion. The age of the SNR computed here is 9.0$\times$10$^{3}$~yr (if the SNR is in its Sedov-Taylor phase) or  9.2$\times$10$^{3}$~yr (if the SNR is in its radiative phase) and is in agreement with  the age estimated  by other authors, showing that this SNR is still young.  The initial energy deposited in the ISM by the SN explosion derived in this work is in the range values of E$_0$=1.8 to 5.2$\times$10$^{50}$~erg (either the SNR is in its Sedov-Taylor or radiative phase of evolution).  These values are not so different from the energy deposited by a supernova forming a pulsar. This suggests that it is not possible to identify supernova forming magnetars from supernova forming typical pulsars from the energies deposited in the ISM.

\section*{Acknowledgements}
We are grateful for financial support from CONACyT-253085. This work was also supported by DGAPA-UNAM PAPIIT-IN103116 and PAI-UAEH-2016-3280. 
I.F-C. acknowledges IPN-SIP project 20171612.
M.S.C. acknowledges CONACyT for a doctoral scholarship.

We acknowledge the very positive input of the anonymous referee that helped in improve the article.

Based upon observations carried out at the Observatorio Astron\'omico Nacional on the Sierra San Pedro M\'artir (OAN-SPM), Baja California, M\'exico.

%%%%%%%%%%%%%%%%%%%%%%%%%%%%%%%%%%%%%%%%%%%%%%%%%%



%\bibliographystyle{mnras}


%%%%%%%%%%%%%%%%%%%%%%%%%%%%%%%%%%%%%%%%%%%%%%%%%%

%%%%%%%%%%%%%%%%% APPENDICES %%%%%%%%%%%%%%%%%%%%%

%\appendix



%%%%%%%%%%%%%%%%%%%%%%%%%%%%%%%%%%%%%%%%%%%%%%%%%%

% Don't change these lines
\bsp	% typesetting comment
\label{lastpage}
\end{document}